# Dynamical vortex phase diagram of 2D superconductivity in gated MoS$_2$

Yu Saito[1,2]*, Yuki M. Itahashi[1], Tsutomu Nojima[3] and Yoshihiro Iwasa[1,4]*

[1] *Quantum-Phase Electronics Center (QPEC) and Department of Applied Physics, The University of Tokyo, Tokyo 113-8656, Japan*

[2] *California NanoSystems Institute, University of California at Santa Barbara, Santa Barbara CA 93106, USA*

[3] *Institute for Materials Research, Tohoku University, Sendai 980-8577, Japan*

[4] *RIKEN Center for Emergent Matter Science (CEMS), Wako 351-0198, Japan*

*Corresponding author: yusaito@ucsb.edu, iwasa@ap.t.u-tokyo.ac.jp

**ABSTRACT**

Recent discoveries of two-dimensional (2D) superconductors have uncovered various new aspects of physical properties including vortex matter. In this paper, we report transport properties and a dynamical phase diagram at zero magnetic field in ion-gated MoS$_2$. In addition to the universal jump in the current-voltage characteristic showing unambiguous evidence of the Berezinskii-Kosterlitz-Thouless (BKT) transition, we observed multiple peaks in the temperature- and current-derivative of the electrical resistance, based on which a dynamical phase diagram in the current-temperature plane was constructed. We found current-induced dynamical states of vortex-antivortex pairs, containing that with the phase slip line. Also, we present a global phase diagram of vortices in gated MoS$_2$ which captures the nature of vortex matter of clean 2D superconductors.

## I. INTRODUCTION

Superconductivity in 2D materials [1] not only expands a new materials paradigm [2–9], but also offers opportunities to explore various novel phenomena of 2D superconductors, which were masked by disorder or randomness in conventional metallic thin film superconductors with amorphous or granular structures. Examples include the quantum metallic state (a vortex liquid state due to quantum fluctuation) as a ground state [10–12], the quantum Griffiths state [13–16], enhanced in-plane upper critical field [17–19], nonreciprocal superconductivity [20–22] and unusual behavior of superconducting gap against magnetic field $B$ [23,24]. Such series of discoveries originating from the minimal disorder and the noncentrosymmetricity accompanying high crystallinity suggest that emerging 2D superconductors form a materials platform for new physics of superconductivity.

While the quantum phases [10–16,25] and the vortex dynamics [26–31] under magnetic fields have been intensively studied in 2D superconductors, the dynamical properties at high current and zero magnetic field have been largely unexplored. It has been well known that in 2D superconductors the zero-ohmic-resistance state is achieved by the Berezinskii-Kosterlitz-Thouless (BKT) transition [32–34]. Above the BKT transition temperature ($T_{BKT}$), the spatial phase fluctuation of the order-parameter causes the thermally-excited unbinding of vortices and antivortices (V-AV), which can move freely and result in the ohmic dissipation. When such vortices are bound in the form of V-AV pairs, the BKT transition to the zero-ohmic-resistance state occurs at a temperature of $T = T_{BKT}$ in the low current limit. Experimentally, the existence of the BKT transition has been confirmed by the universal jump of the power α in the current ($I$) -voltage ($V$) characteristics ($V \propto I^\alpha$) at $T_{BKT}$ and by the scaling behavior of linear resistance ($R$) obeying the Halperin-Nelson equation in the critical region just above $T_{BKT}$ [35,36]. These transition features have been discussed in various systems such as disordered 2D superconductors [37–39] and high-$T_c$ superconductors [40]. On the other hand, the zero-field V-AV dynamics beyond the low current region in 2D systems, which would lead to understanding how the BKT state transfers to the normal state by increasing the current, have remained unsolved yet.

The dynamics of vortices, whether in zero or finite magnetic fields, has been a long-standing research field from the viewpoint of fundamental interests in interacting particles as well as application of superconductors at high current density. While the dynamics of slow Abrikosov vortices in magnetic field ($B$), containing the plastic flow state and the moving lattice state, has been thoroughly investigated in various 2D superconductors [26–28], there are still many ongoing discussions on the dynamics of ultrafast vortices under high current density [31,41]. One of the typical examples is the sudden jump of voltage induced by the vortex motion as observed in various superconductor films [29,30,42–44], which was interpreted based on the flux-flow instability predicted by Larkin and Ovchinnikov in the early stages [45] and was recognized later as the possible appearance of the phase slip lines [46], namely, flow channels of the Josephson-like (kinematic) vortices [47–49]. While such fast dynamical states of vortices were discussed, in many cases, with the dirty systems under $B$ [30,48–50], the experiments in 2D superconductors under zero magnetic field, which indeed corresponds to the original condition of the theoretical prediction, is few [51]. In addition, whether the phase slip lines appear even in clean systems is unclear. In such a situation, exploring the dynamics of V-AV in clean 2D superconductors with weak pinning at zero magnetic field is highly desirable. Indeed, a discussion on dynamical behavior has been recently reported on a few layer $NbSe_2$ flakes and some data are interpreted using a model based on the phase slip phenomena [52].

Here, we report on the current-induced dynamical states of V-AV pairs at zero magnetic field in a gate-induced 2D superconductor, $MoS_2$, through the measurements of transport properties as a function of temperature and current. In addition to confirming the standard BKT transition in low current and high temperature, we observed the anomalous transport with the non-monotonic increase against temperature and current in the high current and low temperature region. The present results suggest the existence of multiple dynamic vortex phases between the BKT state and the normal state at zero magnetic field. Including the present result, we propose a global magnetic field-current-temperature phase diagram, which offers a comprehensive understanding of vortex matter in clean 2D superconductors with weak pinning potentials.

## II. RESULTS

### A. Transport properties in ion-gated MoS$_2$

We prepared thin flakes of MoS$_2$, which is a typical layered semiconductor, from bulk single crystals by mechanical exfoliation, and patterned Au (90 nm)/Cr (5 nm) electrodes onto an isolated thin flake in a standard four terminal and Hall bar configuration with a side gate electrode on Si/SiO$_2$ substrate. To complete an electric-double-layer transistor (EDLT) structure (Fig. 1a), we put a drop of ionic liquid, N,N-diethyl-N-(2-methoxyethyl)-N-methylammonium bis (trifluoromethylsulphonyl) imide (DEME-TFSI) on the above device, as a gate medium. Here, the channel length $L$ and width $W$ are 2.0 μm and 4.0 μm, respectively. By an application of a gate voltage of 4.5 V at 230 K, the MoS$_2$-EDLT showed the metallic $T$ dependence of sheet resistance $R_{sheet}$ (d$R_{sheet}$/d$T$ > 0), followed by the superconducting transition, as shown in Fig. 1b. Here, the current for the $R_{sheet}(T)$ measurement is 5 μA. The residual-resistance ratio (RRR) of the sample, which is defined R(200K)/R(10K) here since the gate is applied 230 K and ionic liquid will be frozen around 200K, is roughly 7~8. This value is indeed comparable to the RRR (R(200K)/R(10K)) of a bilayer (7~8) and monolayer NbSe$_2$ (8~9) [19] and much higher than the RRR of ~1.5 in CVD grown [53] and MBE-grown [54] NbSe$_2$ films, which shows high crystallinity of the present sample. To determine the critical temperature, we used Aslamazov-Larkin (AL) [55] and Maki-Thompson (MT) model [56,57], where the excess sheet conductance by the fluctuation ΔG is given by: $\Delta G = \frac{e^2}{16\hbar}\left(\frac{T_{c0}}{T-T_{c0}}\right) + \frac{e^2}{8\hbar}\frac{T_{c0}}{T(1-\delta)-T_{c0}}\ln\left(\frac{T-T_{c0}}{\delta T}\right)$, where $T_{c0}$ = 7.6 K (mean-field transition temperature) and δ = 0.06 (pair-breaking parameter), used as fitting parameters with $e$ the elementary charge and $\hbar$ the Dirac's constant (the inset of Fig. 1b). The sheet carrier density $n_s$ of this device was 1.19 × $10^{14}$ cm$^{-2}$, which was confirmed by the Hall measurements at 15 K. According to the previous studies [17,18], the application of such a high gate voltage leads to the accumulation of the carriers in the topmost layer with effective thickness $d$ ~ 1 nm. It is noted that the normal state sheet resistance $R_N$ is around 220 Ω, which results in $k_F l = \frac{1}{ss'}\frac{2h}{e^2}\frac{1}{R_N} \sim 59$. This value is much larger than the Ioffe-Regel limit ($k_F l$ ~ 1). Here, $k_F$ is the Fermi wave number, $l$ is the mean free path and $s$ and $s'$ are spin and valley degree of freedom, respectively. In addition, in the superconducting state, the value $l/\xi_0$ is about 0.37, where $l$ and $\xi_0$ (Pippard's coherence length) are 30 nm and 81 nm, respectively. All the properties mentioned above suggest the 2D superconductor on the surface of MoS$_2$ has the less-disordered nature far from dirty limit.

Figure 1c shows the *I-V* characteristics of the ion-gated MoS$_2$ at various temperatures ranging from 2 and 10 K. We found that α, which is the exponent of the relation at low current limit, $V \propto I^\alpha$, shows almost 1 above 6.5 K, gradually increases with decreasing temperature, and then abruptly jumps to 3 at 6.0 K (Fig. 1d). From this data, we determined the BKT transition temperature $T_{BKT}$ = 6.0 K. Furthermore, the plot of ln($R_{sheet}$) as a function of $[(T_{c0}-T)/(T-T_{BKT})]^{1/2}$ at $I$ = 5 μA in the ohmic region shows linear behavior (Fig. 1e), which is consistent with a Halperin-Nelson model $R = R_0 \exp\left\{-2b[(T_{c0}-T)/(T-T_{BKT}))]^{1/2}\right\}$ [35,36,58]. This implies that the decrease in the resistance above $T_{BKT}$ is caused by the development of the V-AV correlation with the characteristic length scale $\xi_+ \sim \xi \exp\left\{b\left[(T_{c0} - T)/(T - T_{BKT})\right]^{1/2}\right\}$, with $b$ numerical constants of order of unity. To obtain the dashed line in Fig. 1e, we used $T_{c0}$ = 7.6 K and $T_{BKT}$ = 6.0 K extracted from the AL and MT fit and $I$-$V$ characteristic, and $R_0$ = 508 Ω and $b$ = 1.06 as the materials-dependent fitting parameters. Both the abrupt change of $\alpha$ in $I$-$V$ characteristic and the scaling of $R_{sheet}$-$T$ dependence is therefore in fair agreement with the standard BKT picture of 2D superconductors.

**B. Anomalous transport at high excitation current**

To investigate how the BKT state transfers to the normal state by the application of the high current, we measured the non-linear sheet resistance $R_{NL}$ at various bias currents ranging from 5 to 200 μA (Fig. 2a). While the $R_{NL}$–$T$ curves strongly depend on $I$ above 5 μA, we note the region with $R_{NL}$ smaller than the noise level (BKT state as discussed later) is maintained until 55 μA in the measured $T$ range > 2 K. This feature shows marked contrast to the case of application of out-of-plane magnetic fields, in which the BKT state is very fragile even in a very small $B$ [10,16]. More importantly, we found that the anomalous step-like behavior in $R_{NL}$-$T$ curves appears at middle bias currents ($I$ = 35 – 65 μA), followed by the $T$-independent behavior at high bias currents ($I$ > 70 μA). On the other hand, such step-like behavior becomes obscure below $I$ = 30 μA, suggestive of a single transition at low enough currents ($I$ < 20 μA). In order to quantify such features, we plot the temperature derivative of resistance d$R_{NL}$/d$T$ at various currents as shown in Fig. 2b. We found the two characteristic temperatures $T_1$ (white triangles) and $T_2$ (black triangles) with the peaks, corresponding to the inflection points in the $R_{NL}$-$T$ curves denoted by white and black circles in Fig. 2a. So far, it has not been discussed intentionally how the BKT state below $T_{BKT}$ or the flux-flow state above $T_{BKT}$ (< $T_{c0}$) are extended along the current axis, although one notices that the dynamics of V-AV in the both states plays a central role there [35,39]. Figure 2b suggests the existence of intermediate dynamical states of V-AVs between the BKT and the fluctuation region of the order parameter amplitude with clear boundaries.

For further understanding of the current-induced anomalous behavior observed in Fig. 2, we extracted the $d\ln V/d\ln I$ against $I$ from the data in Fig. 1c. Figs. 3a, b and c show $d\ln V/d\ln I$ as a function of $I$ at various temperatures, for the whole regime (0 < $I$ < 100 μA) (a) and magnifications around 50-76 μA (b) and 45-95 μA (c), respectively, which represent the changes in the nonlinear behavior between the BKT state and fluctuation/normal state. Here, we provide the data at representative temperatures for clarity/visibility. We found three distinct peaks at $I_{c1}$ (black triangles in Fig. 3a), $I_{c2}$, (white diamonds in Fig. 3b) and $I_{c3}$ (white triangles in Fig. 3c). As displayed in Figs. 3a-c, $I_{c1}$, $I_{c2}$ and $I_{c3}$ shift toward the low current with increasing temperature. At the same time, these peaks become broad and eventually undefined around 6.5, 3.0, and 6.0 K for $I_{c1}$, $I_{c2}$, and $I_{c3}$, respectively. It should be noted that the $I$-$V$ curves keep nonlinearities with $d\ln V/d\ln I > 3$ after the step-like increases of $V$ (corresponding the peaks of $d\ln V/d\ln I$ in Fig.3) at $I_{c1}$ and $I_{c2}$ below $T_{BKT}$, implying these two anomalies are completely different from the thermal quench process to the normal state. Also, as shown in Fig. S1, the peaks of $I_{c2}$, and $I_{c3}$ becomes vanishingly small or washed out at 0.2 T (corresponding to the 10% of perpendicular upper critical field (~2 T). These data suggest that there are three dynamical transitions or crossovers between the BKT (superconducting) state and the amplitude fluctuation (partly normal) state at low temperature. Because the dissipation of the superconductor at low temperature originates from the motion of vortices, especially that of V-AV pairs in zero magnetic field, the anomalies are attributed to the abrupt change in the velocity of dissociated V-AV pairs or to the transition from the V-AV flow state to the normal or fluctuation (partly normal) regime, as discussed later.

Such step-like behavior of the $I$-$V$ curve has been reported in a 100 nm thick Sn film in a narrow temperature range of 10-30 mK below $T_{c0}$ [51] and in a wider temperature range in few layer NbSe$_2$ flakes [52], which is ascribed to the occurrence of the phase slip line(s) with kinematic vortices running across the samples plane [30,48–50]. This phenomenon seems closely related to our observation as discussed later. In most 2D superconductors, the abrupt jump of *I-V* characteristic in the BKT state results in reaching the normal state with ohmic behavior or the destruction of superconductivity [38,59]. One plausible reason for the successful observation of the multiple nonequilibrium states in this work is because the present 2D system belongs to relatively clean systems with weak pinning based on the high crystallinity, which expands the *I-V* measurement window. In the case of superconductors with strong pinning, the pinning effect should decrease the measured voltage, blurring the subtle dynamical changes. In addition, if the system is in the dirty limit, the fast motion of unbound V-AV pairs, originating from the high flux flow resistance, narrows the region where the system shows nonlinear dynamic behavior against current. The present system with less disorder and minimal pinning potential, does not suffer from such problems and thus provides an ideal playground for the V-AV dynamics, which makes an easier access to the intrinsic nonequilibrium state of 2D superconductors.

## III. DISCUSSION

### A. Current-temperature phase diagram

Figure 4 provides a current-temperature (*I* - *T*) phase diagram at zero magnetic field in ionic-gated MoS$_2$, combined with a color map of d$R_{NL}$/d$T$ values as a function *I* and *T* in the measurement regime of *T* > 2 K and *I* > 5 μA. Light blue, red and white circles show $I_{c1}$, $I_{c2}$ and $I_{c3}$, respectively. Red and white triangles show peaks at lower ($T_1$) and higher ($T_2$) characteristic temperatures in Fig. 2, respectively. Green triangles, which are defined by the temperatures where the resistance becomes 95% of the normal state resistance ($T_3$) in Fig. 2a, shows an onset of amplitude fluctuation of order parameter. In Fig. 4, we note $T_2$ merges with $I_{c3}$, constructing the same boundary. On the other hand, $T_1$ almost coincides with $I_{c1}$ at high temperature, but separates from it at low temperature below ~3.5 K, forming an additional boundary together with $I_{c2}$.

Below $I_{c1}$, the system maintains the BKT state (Region A) or the BKT critical state with the finite V-AV correlation length $\xi_+$ (Region B). According to the BKT scenario, the electrical resistance is caused by flows of the unbound V-AV pairs [35,36]. In Region A, such free vortices (dissociated V-AVs) are induced by applied current and the density of free vortices $n_f$ is described as

$$n_f = C \cdot I^{\alpha - 1}, \tag{1}$$

with $C$ the temperature dependent numerical constant. This relation is derived from the dissociation probability of the V-AV pairs following $\Gamma \propto n_f^2 \propto \exp[-U(r_c)/k_B T]$, where the energy barrier $U(r_c) = Kd \ln(r_c/\xi)$ and the threshold length for dissociation $r_c = KS/(I\phi_0)$ are determined by the $r$ that maximizes the sum of the energy loss of V-AV interaction $Kd \ln(r/\xi)$ and the energy gain owing to the Lorentz force $I\phi_0 dr/S$ with the distance $r$ of the thermally excited V-AV pair. Here, $K = \frac{1}{\mu_0}\left(\frac{\phi_0^2}{4\pi\lambda^2}\right)$, $\phi_0$ the flux quantum, $\xi$ the GL coherence length, $\lambda$ the magnetic field penetration depth, $k_B$ the Boltzmann constant, $d$ (= 1.0 nm) the effective thickness, $S$ (= $d \times W$ = 1.0 nm × 4.0 μm) the cross section of current flow and $\alpha = Kd/2k_B T + 1$. By using the Eq. (1) and the vortex velocity $v_f = \phi_0 I/\eta S$, which is determined by the balance of the viscous drag force $\eta v_f d$ and the Lorentz force $\phi_0 Id/\eta S$ working on the vortices, we obtain the *I-V* characteristic as

$$V = (n_f \phi_0 v_f)L = \frac{n_f \phi_0^2 I}{\eta}\left(\frac{L}{S}\right) = C\left(\frac{\phi_0^2 L}{\eta S}\right)I^\alpha \tag{2}$$

with $\eta$ the vortex flow viscosity and $L$ (= 2.0 μm) the distance between the voltage contacts, which is consistent with the experimental data in Region A. In Region B, on the other hand, the resistance in the low current limit is ohmic because of the thermally-dissociated spontaneous (uncorrelated) V-AV pairs with the large distance $r_c > \xi_+$, but the non-ohmic component appears at high current because of the unbinding of correlated V-AV with $r_c < \xi_+$, leading to additional dissipation according to the Eq.(2).

As *I* increases from Regions A or B, the *I-V* characteristics show diverging behavior at $I_{c1}$, corresponding to the peak of dln*V*/dln*I*, and then transfer to Region C or D, depending on temperature. This diverging behavior reminds us of the occurrence of the flux flow instability [45], where $\eta$ at high current (at high $v_f$) is modified as $\eta_0/\{1 + (v_f/v_0)^2\}$, where $\eta_0$ is viscosity at low current limit and $v_0$ the critical velocity. In such a case, it is expected that the *I-V* curves jump to the different branches with the much smaller vortex flow viscosity (much faster vortex velocity) at $v_f = v_0$. Within the framework of existing theories, the formation of the phase slip line with kinematic vortices may be a leading candidate for the anomaly at $I_{c1}$. Another possibility is the failure of the thermally activated process in Eq. (1) because of the decrease in $U(r_c)$ with increasing *I* to the same order of $k_B T$, which will lead to the abrupt increase in $n_f$ in Eq. (2). It is noted that because the region B only exists in the high temperature and low current region, the boundary between B and C disappear at low temperature and high current. Also, with the appearance of Region D at low temperature and high current, two boundaries of A and D and of C and D appear, branching from the A-C boundary.

In order to understand the emergence of Region C, we analyzed the *I-V* curves from 4.5 to 5.8 K in detail (Fig. 5a), for which a direct transition from Region A to C is observed and the data show clearly the BKT-type power law at low voltage limit explained by Eq. (2). As shown in Fig. 5a, we note the slopes of *I-V* curves in log-log plots do not change so much by comparing the data before and after the step-like increase at $I_{c1}$ (black dashed lines in Fig. 5a). This implies that the nucleation process of dissociated V-AVs, namely the description of $n_f$ in Eq. (1), is almost unchanged at least around the jump in *V*. Then, the transition from Region A to C is ascribed to the abrupt increase in $v_f$ because of the decrease in $\eta$ in Eq. (2) rather than the increase in $n_f$. Indeed, the calculated $U(r_c) = Kd \ln(r_c/\xi) = Kd \ln(KS/\xi\phi_0 I)$ with $\xi = 12.0$ nm and $\lambda = 205$ nm [16,18] is 5 -15 meV at $I_{c1}$, much larger than $k_B T$, meaning that the thermally activated process (Eq, (1)) hold at least around $I_{c1}$. Thus, the phase slip line with kinematic vortices likely occurs in Region C.

To check this scenario, we directly calculated $v_f$ from the data by using the relation

$$v_f = \frac{V/L}{n_f \phi_0} = \frac{V/L}{CI^{\alpha-1}\phi_0}, \tag{3}$$

assuming that Eq. (1) hold before and after the transition at $I_{c1}$. Here, $C$ and $\alpha$ are obtained by fitting the linear part of $\ln I$ - $\ln V$ data in the low current region below $I_{c1}$ (typically around 10 - 40 μA) in Fig. 5a with Eq. (2) and $\eta = \eta_0 = \frac{\phi_0^2 d}{2\pi\xi^2 R_N}$ according to the Bardeen-Stephen law. Figure 5b shows $v_f$ as a function of $I$ between 4.5 and 5.8 K, where the black triangles in the figure display the position of $I_{c1}$. As can be seen in Fig. 5b, $v_f \approx 10^3$ m/s below $I_{c1}$ which is the typical velocity of Abrikosov vortex [60], abruptly increases to $v_f \approx 10^5 - 10^6$ m/s above $I_{c1}$ and below 5 K. This value is indeed consistent with the picture of kinematic vortices previously reported [51]. We note that the decrease in $v_f$ with further increasing current is unphysical. This may originate from the failure of the assumption of Eq. (1) in Region C because of the interaction between the kinematic vortices, which limits the actual value of $n_f$ and thus leads to underestimating $v_f$ at high current region. With further increasing current in Region C, the order parameter of phase slip line decreases and becomes vanishingly small. Once the current increase over the $I_{c3}$, the flux flow state becomes partially normal and then the system goes to the fluctuation state (Region E; $T_2 < T < T_3$, $I > I_{c3}$). The peak of d$\ln V$/d$\ln I$ at $I_{c3}$ may correspond to disappearance of the vortex picture. It is noted that in the geometry of EDLT, the gate electric field is not applied under electrodes and thus those regions are insulating at low temperatures, which might lead to the focused current flow between the Hall electrodes. Such a inhomogeneous current flow might assist the formation of the phase slip line and should be taken into account, as discussed in the previous work on $NbSe_2$ [52]. To confirm such inhomogeneity of the supercurrent and clarify microscopic origin of the phase slip line needs more direct probe such as nanoSQUID-on-tip and thus is left for the future study.

In the dynamical phase diagram of Fig. 4, the appearance of Region D below the boundary of $I_{c2}$ and $T_1$ at low temperature lower than 3.5 K is also quite unexpected and its explanation remains an open question. It has been recognized that there are two types of the dissipative states due to fast vortex motion; one is the phase slip line with kinematic vortices as discussed above and the other is vortex street with the fast flow of Abrikosov vortices, which is fast as compared to the homogeneous flux flow, but not too fast as the kinematic vortices [46,49]. Recently, the observation of the fast vortex flow channels, where the vortices keep the nature of the Abrikosov vortices, has been reported through the scanning nano-SQUID imaging technique [41]. Although our experiments were performed at zero magnetic field, a similar state may occur in Region D. As another candidate, the appearance of multiple phase slip lines might be possible. In this case, $I_{c1}$ and $I_{c2}$, which correspond to the onset of the first and the second phase slip line, respectively, should be observed separately for all the temperature region [52], which may not be the case in our experiments.

It is emphasized that we can rule out the possibility of the spatial nonuniformity of the sample as the origin of the multiple peaks in $d\ln V/d\ln I$ and $dR_{NL}/dT$ because those $I_{c1}$, $I_{c2}$ and $I_{c3}$ (or $T_1$ and $T_2$) show the different temperature (or current) dependence with each other. Especially, the $T_1$ line (or $I_{c2}$ line) is merged into the $I_{c1}$ line at high temperature in Fig. 4, which results in the limited observation of step-like features in $R_{NL}(T)$ in Fig. 2 only between 35 and 65 μA with the absence in smaller current region. In addition, even the $I_{c1}$ line is connected to the $T_2$ line near $T_{BKT}$. In the case of inhomogeneous superconductors, such anomalies should appear rather in a wide range of temperature and current with scaled behavior. Furthermore, we found that in the normalized plot (Fig. S2), $I_{c3}(T_2)$ deviates from $I_{c1}$ or $I_{c2}(T_1)$ except for the temperature region near $T_{c0}$. This implies that $I_{c3}(T_2)$ at low temperature has a different origin from $I_{c1}$ and $I_{c2}(T_1)$ at least. It is unexpected that $T_3$ shows a similar scaled curve to $I_{c1}$ and $I_{c2}(T_1)$. Although the origin of their similarity is unclear, this behavior is far from that for the depairing current scaled as $(1-T/T_{c0})^{2/3}$ with a positive curvature. We also note that the temperature dependence of the normalized of $I_{c1}$ or $I_{c2}$ is very similar to that of the normalized critical current in 2D $NbSe_2$ [52] with a negative curvature, which may be common properties of the characteristic current for the phase slip line in the clean superconductors, while the $I_{c1}$ at the high temperature shows more gradual slope than the normalized critical currents of 2D $NbSe_2$.

In the present study the Joule heating is fairly excluded because the dissipation power $P = IV$ is the order of 1-100 nW (Fig. 6), which is small by a factor of more than $10^3$ compared to the previous studies [30,61]. In addition, the temperature dependence of $P$ at $I_{c1}$ and $I_{c3}$ in Fig. 6 is inconsistent with the Joule heating model, because $P$ should decrease with increasing temperature and goes to zero with approaching $T_c$.

**B. Magnetic field-current-temperature phase diagram**

Finally, we discuss a *B-I-T* phase diagram of gate-induced superconductivity in $MoS_2$ which represents single-crystal-based clean 2D superconductors with weak pinning potentials. Figure 7 shows a schematic global phase diagram based on Fig. 4, incorporating the *B-T* phase diagram at zero current limit [16] and the *B-I* nonequilibrium phase diagram at low *T* [21]. The zero resistance in the low current limit is achieved by the BKT state at zero magnetic field, and pinned vortex states at small magnetic field (ZR in Fig. 7). When the magnetic field is increased below $T_{BKT}$ at low current, there appears a "quantum metallic state" with *T*-independent resistance as reported in Refs. [10,11,16], which corresponds to a vortex liquid state as a ground state. In contrast to the conventional vortex liquid in quasi-2D bulk superconductors [62–65], the liquid state in the present system survives down to $T = 0$, and consequently, the zero resistance state under magnetic field (vortex solid state) is very fragile. The origin of this state is still in debate [12] but the most plausible model for the finite ohmic resistance, in other words, the energy dissipation, is the quantum collective creep of vortices due to the quantum fluctuation and weak pinning [10,16]. This quantum metallic state crossovers to thermally activated creep state at higher temperature regions, and to the free vortex flow state at higher magnetic field.

Another notable state is seen in the high magnetic field/low temperature region at the zero current limit. Just before the superconductivity is completely destroyed by the application of magnetic field passing through the vortex flow state, the system becomes highly nonuniform, where the superconducting puddles (rare region) exists in the sea of normal state with a rather long time-scale and finite length scale due to the quantum fluctuation. This is called the quantum Griffiths state [13,16], where the slight residual disorder plays an central role. It is interesting that a phenomenon controlled by disorder appears in relatively clean systems, not in the conventional metallic thin film superconductors with full of disorder. This is possibly because subtle phases like the quantum Griffiths state is masked in highly disordered systems.

In the $B$-$I$ plane at the lowest temperature, the zero resistance state and the quantum metallic state in a magnetic field are rapidly suppressed with increasing the current [21]. In particular, $MoS_2$ has an in-plane broken inversion symmetry, and therefore the nonreciprocal response is observed under out-of-plane magnetic field. This allows us to distinguish the quantum metallic state for $B < B_1$ and the classical vortex flow (or plastic flow) for $B_1 < B < B_2$ due to vortex ratchet effect. The two boundaries $B_1$ and $B_2$ seemingly merge in the zero-current limit and thus, the quantum metallic state is dominating in the thermally equilibrium $B$-$T$ plane [21].

The main subject of the present work was on the $I$-$T$ plane, which can be reasonably connected to the $B$-$T$ and the $B$-$I$ plane to form a global phase diagram as shown in Fig. 7. The dynamics of rapidly moving vortices is attracting recent interest [31], since it is related to general questions on the stability of the topological defects. In this context, it is interesting to revisit the stability of the BKT state with current induced V-AV. Researchers have shown that the BKT state is destroyed and the fluctuation region or normal state is immediately recovered in conventional 2D metallic thin film superconductors. However, the present results on gated MoS$_2$ revealed that the transition from the BKT state to the normal or fluctuation state is much more complex, exhibiting at least three intermediate dynamical states, B, C, and D in Fig. 4. We propose interpretations of the B and C states in terms of the thermally dissociated V-AV pairs and the kinematic vortex state with the phase slip line, respectively. D might be understood by the vortex street (which may develop into the plastic flow in a magnetic field).

## IV. CONCLUSION

In conclusion, we investigate the dynamical behavior of vortices and antivortices at zero magnetic field in a 2D superconductor, ion-gated MoS$_2$. After confirming the BKT transition by the universal jump of the power α in the $I$-$V$ characteristics ($V \propto I^\alpha$) at $T_{\text{BKT}}$, we established a dynamical phase diagram on the $I$-$T$ plane. We proposed a scenario of the occurrence of ultrafast flow of vortices, which may be regarded as the kinematic vortices for a dynamic state based on the flux flow instability of the current driven vortices and antivortices. Furthermore, we present a comprehensive phase diagram including nonequilibrium states on the $B$-$I$-$T$ space. This phase diagram offers a global view on how the superconducting (BKT) state evolves to the normal state though multiple intermediate states by $B$, $I$, and $T$, and have never been observed in conventional disordered 2D superconductors, which are described by the dirty boson picture [25]. We suggest that the unexpected richness of the dynamical vortex diagram is attributed to the combination of the weak pinning force, the small resistance, and enhanced fluctuations at the 2D limit.

Nonlinear properties of superconductivity are rarely observed due to Joule heating of samples and poor heat transfer between samples and substrates. This leads to a sharp jump from the BKT state directly to the normal state in a current-voltage characteristic without any intermediate states (i.e., the thermal instability). In marked contrast, the ionic-gating induces superconductivity on a crystal surface, and thus the rest of the crystal may serve as a good heat bath, not only reducing thermal instability, but also making the dynamic properties at high currents accessible. The present results highlight the uniqueness and importance of the gate-induced 2D superconductivity, offering a challenge to construct a theoretical scheme for clean 2D superconductors.

**V. OUTLOOK**

Recently emerging 2D crystalline superconductors have been offering various opportunities for exploring transport properties reflecting intrinsic quantum phases and nonequilibrium vortex dynamics as well as crystal symmetry driven phenomena, which have been not found in disordered 2D superconductors. One possible direction for future discoveries is to investigate the real space microscopic states by the local microscopy such as STM, ARPES and nanoSQUID as exemplified by the previous study [41]. Another direction is various applications, for example, for the superconducting transistors and superconducting quantum bit. In any cases, the further developments for the fabrication and characterization of 2D crystalline superconductors would be in high demand.

**ACKNOWLEDGEMENTS**

We thank N. Nagaosa for fruitful discussions. Y.S. was supported by the Japan Society for the Promotion of Science (JSPS) through a research fellowship for young scientists (Grant-in-Aid for JSPS Research Fellow, JSPS KAKENHI Grant Number JP15J07681) and the Elings Prize Fellowship from University of California at Santa Barbara. This work was supported by Grant-in-Aid for Scientific Research S (JSPS KAKENHI Grant Number JP19H05602) from JSPS and for Scientific Research on Innovative Areas (JSPS KAKENHI Grant Number JP16H01061) from JSPS.

**AUTHOR CONTRIBUTION**

Y.S., T.N. and Y.I. conceived the idea and wrote the manuscript. Y.S. and Y.M.I. fabricated the devices and analyzed the data. Y.S. designed the experiments and conducted the transport measurements. All authors discussed the results.

## Figure Captions

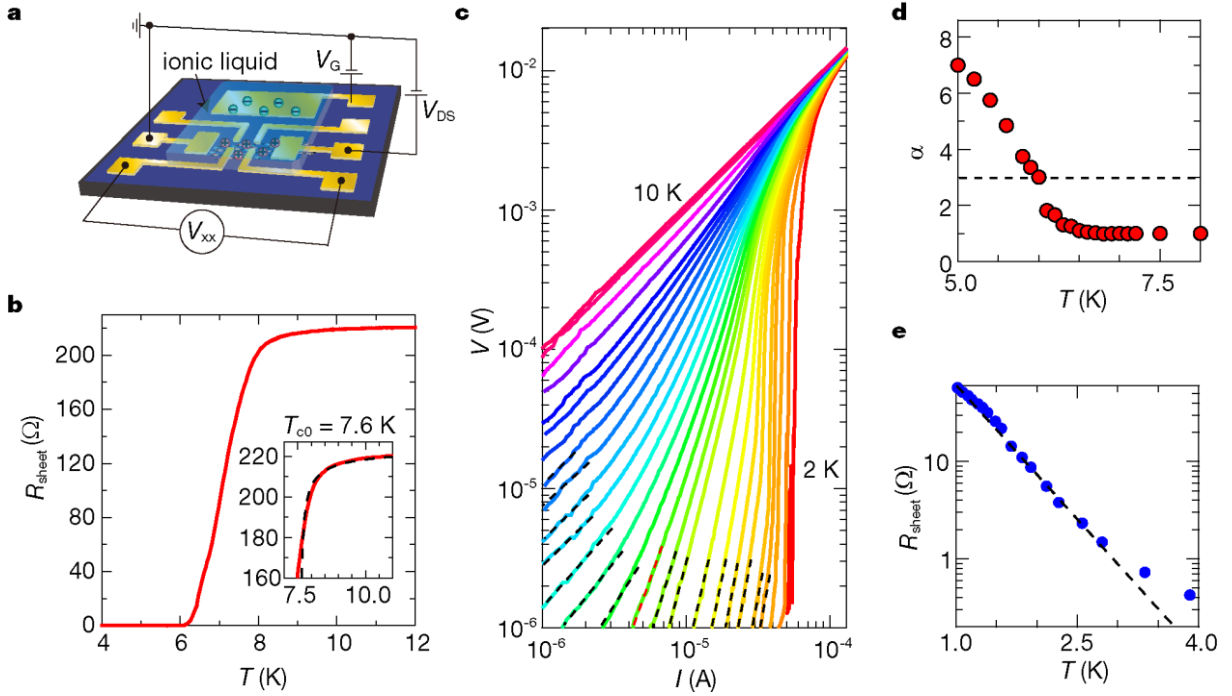

**Figure 1. Berezinskii-Kosterlitz-Thouless (BKT) transition in ion-gated MoS$_2$. a**, A schematic image of electric-double-layer transistor structure. **b**, Superconducting transition in ion-gated MoS$_2$. Inset: A magnified view near the superconducting fluctuation regime. Black dashed curve shows Aslamazov-Larkin (AL) and Maki-Thompson (MT) fit. The superconducting transition temperature ($T_{c0}$) defined by AL and MT fit is 7.6 K. **c**, Current-voltage curves on a logarithmic scale at various temperatures varying in 1 K steps from 2 to 4 K, 4.2, 4.5, 4.7 K, in 0.2 K steps from 5 to 5.8 K, in 0.1 K steps from 5.9 to 7.2 K, 7.5, 8 and 10 K. The short black dashed lines are the log-log fit to the data in the low current limit (below critical region) at each temperature. Red dashed line shows the same fit at the BKT transition temperature ($T_{BKT}$) **d**, Temperature dependence of the power-law exponent α, as deduced from the fits shown in **c**. From this plot, we defined the $T_{BKT}$ is 6.0 K, where the α discretely becomes 3. **e**, Plots of $R_{sheet}$ as a function of $[(T_{c0}- T)/(T-T_{BKT})]^{1/2}$ at 5 μA. The black dashed line indicates the fitting of the Halperin-Nelson equation, $R = R_0 \exp\left\{-2b\left(\dfrac{T_{c0}-T}{T-T_{BKT}}\right)^{1/2}\right\}$

. Here, $R_0$ and $b$, which are materials dependent parameters, are 508 Ω and 1.06, respectively.

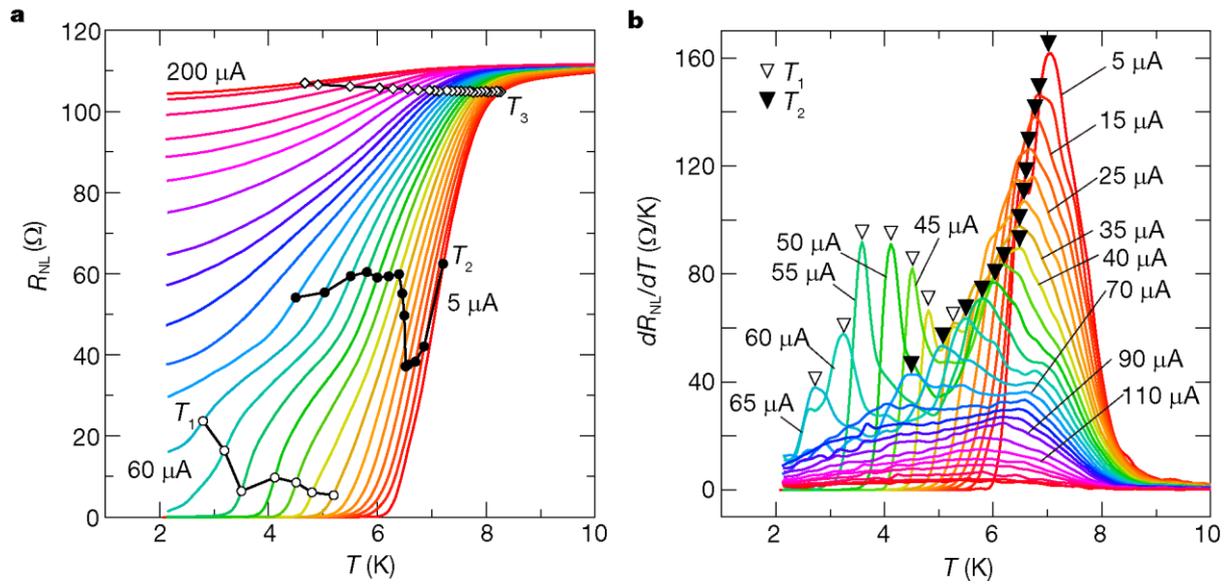

**Figure 2. Longitudinal non-linear resistance $R_{NL}$ and its derivative $dR_{NL}/dT$ versus temperature at various currents. a**, $R_{NL}$ at various bias currents varying in 5 μA steps from 5 and 90 μA, in 20 μA steps from 110 and 170 μA, and 100, 180, 200 μA. The black and white circles show $T_1$ and $T_2$, respectively, which are defined by white and black triangles in Fig. 2b. Also, $T_3$ is defined by the temperatures where the resistance becomes 95% of the normal state resistance, which is indicated by white diamonds. **b**, Temperature derivative of $R_{NL}$ versus temperatures at various currents. Black and white triangles show the double (35-65 μA) and single (5-30 and 70-200 μA) peak positions of $dR_{NL}/dT$ versus $T$ curve, respectively.

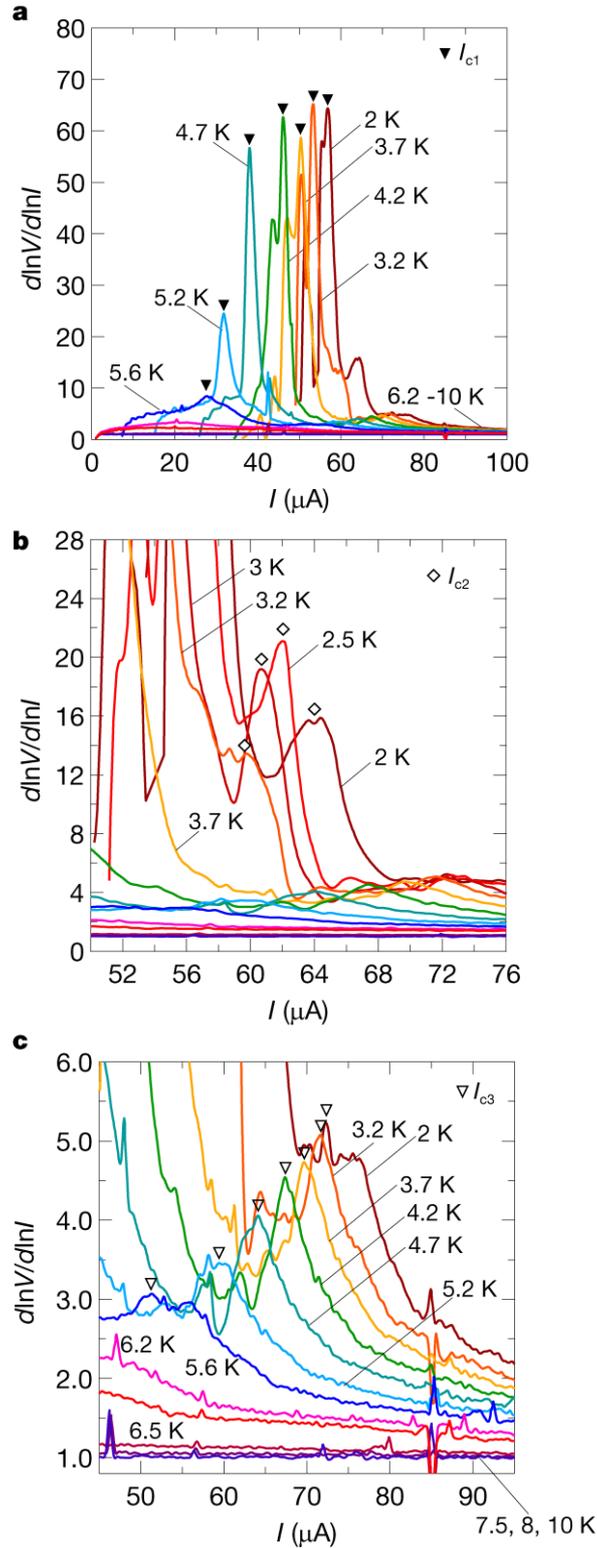

**Figure 3. Differential resistance d ln$V$/d ln$I$ at various temperatures. a**, d ln$V$/d ln$I$ as a function of $I$ at various temperatures from 2 to 10 K. Here, we provided data at representative temperatures. We choose data at 2, 3.2, 3.7, 4.2, 4.7, 5.2, 5.6 and 6.2-10 K. Black triangles show the strongest peak positions below 60 μA. We defined $I_{c1}$ of each d ln$V$/d ln$I$ curve as the current where the dln$V$/dln$I$ curves show these strong peaks indicated by black triangles. **b**, A magnification of Fig. 3a around 50-76 µA. Here, we provide additional data at 2.5 and 3.0 K. White diamonds show the second strong peaks around 50-60 µA. We defined $I_{c2}$ of each dln$V$/dln$I$ curve as the current where the dln$V$/dln$I$ curves show these peaks indicated by white diamonds. **c**, A magnification of Fig. 3a at high current regime. White triangles show the broad peak positions at high current regime above 50 µA. We defined $I_{c3}$ of each dln$V$/dln$I$ curve as the current where the dln$V$/dln$I$ curves show these broad peaks indicated by white triangles.

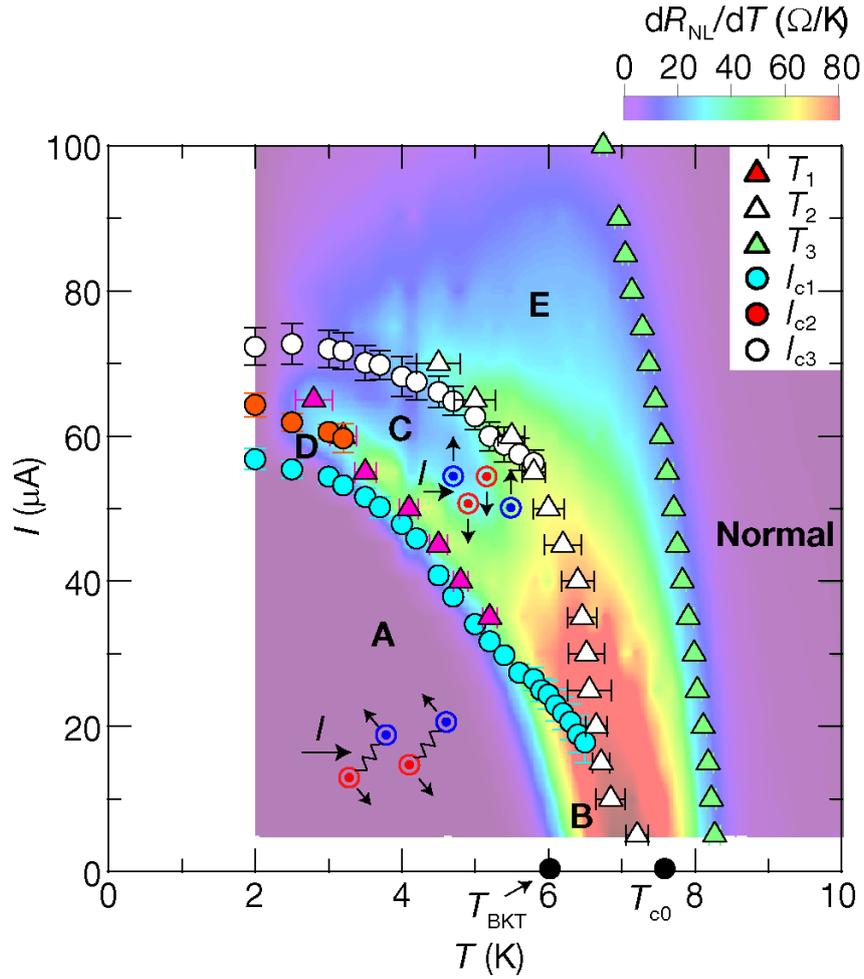

**Figure 4. Current-temperature phase diagram of ion-gated MoS₂ at zero magnetic field.** Schematic images in the regime of the BKT state and vortex flow state show binding vortices/antivortices pairs (Region A), and free vortices and antivortices in the current (Region C). Red and white triangles show peaks of d$R$/d$T$-$T$ curves at lower ($T_1$) and higher ($T_2$) temperatures in Fig. 2b, respectively. Green triangles, which are defined by the temperatures where the resistance becomes 95% of the normal state resistance, shows a boundary between fluctuation regime and normal state. Light blue, red and white circles show $I_{c1}$, $I_{c2}$ and $I_{c3}$, respectively. The vertical error bars for $I_{c1}$, $I_{c2}$ and $I_{c3}$ and the horizontal error bars for $T_1$, $T_2$ and $T_3$ represent uncertainties in determining each data point. The color map of d$R_{NL}$/d$T$ values as a function $I$ and $T$ is displayed in the measurement regime of $T > 2$ K and $I > 5$μA. $T_{c0}$ and $T_{BKT}$ are determined from AL and MT fit and $I$-$V$ characteristic in Fig. 1, respectively. Region A is the BKT state with zero resistance where the correlated vortices and antivortices are binding. Region B ($I < I_{c1}$, $T_{BKT} < T < T_2$) shows the BKT critical state with the finite V-AV correlation length. Region C ($I_{c2} < I < I_{c3}$, $T_1 < T < T_2$) and show nonactivated flux flow region because of unbinding V-AV pairs, which results in intermediate state in non-equilibrium.

Region D ($I_{c1} < I < I_{c2}$, $T < T_1$) shows another intermediate state possibly because of V-AV lattice, where the dissipation process is based on the motion of dislocation-antidislocation pairs, or the BKT phase with quantum fluctuation. Region E ($T_2 < T < T_3$, $I > I_{c3}$) show the fluctuation state.

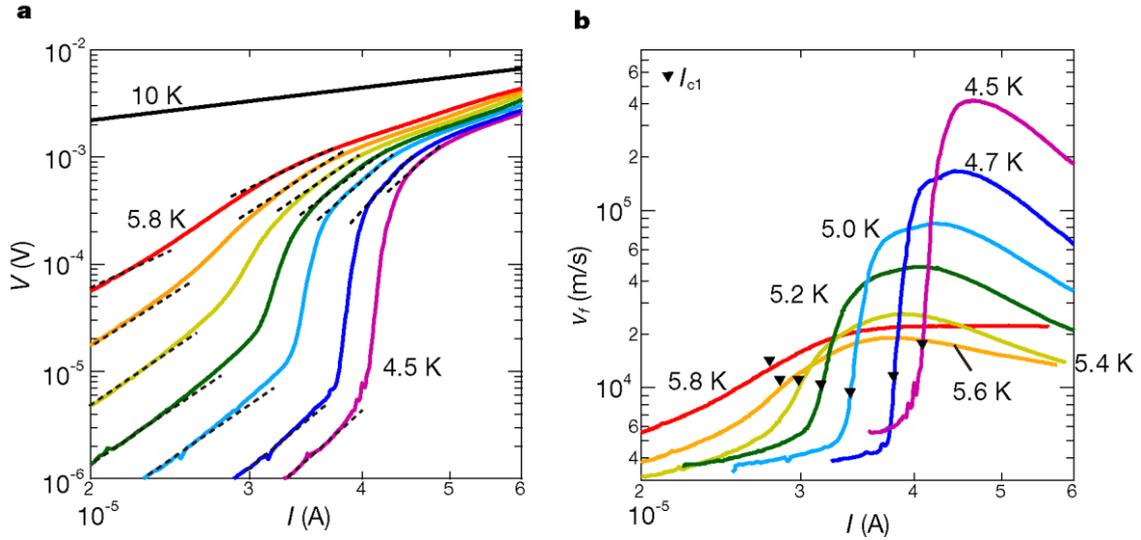

**Figure 5. The velocity of a single vortex versus current. a**, Current-voltage curves on a logarithmic scale at 4.5, 4.7, 5.0, 5.2, 5.4, 5.6, 5.8 and 10 K. Dashed lines show regions following $V \propto I^\alpha$ below and above $I_{c1}$. **b**, Velocity of the single vortex as a function of the current calculated by $v_f = \frac{V/L}{C(T)I^{\alpha(T)-1}\phi_0}$, where $L$ is the length between 4 terminal contact, $\phi_0$ is a magnetic flux quantum. $C(T)$ is a determined the fit of $IV$ curve at the lowest current limit using the relation: $V = 2\pi\xi^2(T)C(T)R_N I^{\alpha(T)}$, where $\xi$ is the GL coherence length $R_N$ is the normal state resistance of 220 Ω. Black triangles show the position of $I_{c1}$.

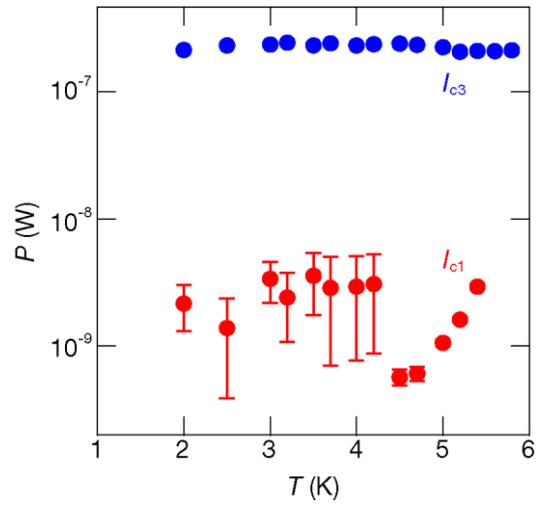

**Figure 6. Dissipated power.** Temperature dependence of the dissipated power at $I_{c1}$ and $I_{c3}$ calculated by $P = IV$, where $I$ is $I_{c1}$(red) or $I_{c3}$ (blue), and $V$ is the values at $I_{c1}$ or $I_{c3}$ at each temperature. The vertical error bars of $P$ at $I_{c1}$ represent uncertainties in determining the voltage at $I_{c1}$ from the measured $I$-$V$ characteristic.

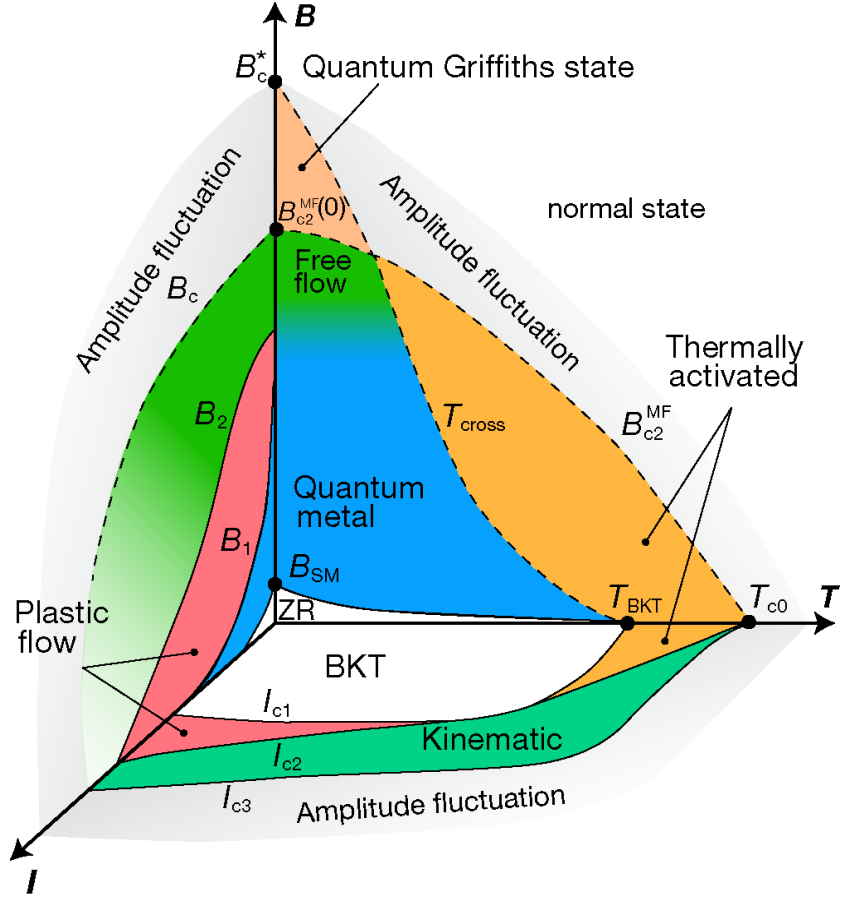

**Figure 7. Comprehensive *B-I-T* phase diagram in single-crystal-based clean 2D superconductors with weak pinning**. The whole phase diagram is based on the results in Fig. 4 and previous works in refs. [16,21]. In the *B-T* plane, white, blue, yellow, green and orange regions show the zero resistance (ZR), the quantum metal, the thermally activated vortex creep, the vortex free flow and the quantum Griffiths states, respectively. Here, $T_{c0}$ is the transition temperature determined by the thermal fluctuation theories (Aslamazov-Larkin and Maki-Thompson model), and $T_{BKT}$ is the BKT transition temperature. $T_{cross}$ is crossover temperature between the thermal activated vortex creep regime and the quantum creep regime, which are determined by the activation plot. $B_{SM}$ is the hypothetical transition magnetic field from the ZR state to the quantum metal (vortex liquid). $B_{c2}^{MF}$ is mean field upper critical field. With increasing magnetic field, the system goes to the quantum Griffiths state up to the characteristic critical magnetic field $B_c^*$ through the free flow state (unpinned vortex). In the *B-I* plane, $B_1$ and $B_2$ are low and high threshold magnetic fields for nonreciprocal response [21], and $B_c$ is the crossover magnetic field between free flow states and fluctuation regime. In the *I-T* plane, white, yellow, green, and red regions correspond to A, B, C, and D states in Fig. 4, respectively. A is the BKT state, and B, C, and D are attributed to the thermally activated V-AV, the kinematic

vortex states, and the vortex street (plastic flow), respectively. Here, $I_{c1}$, $I_{c2}$ and $I_{c3}$ are determined in Fig. 3. Outermost region of these states is governed by the amplitude fluctuation.